\documentclass[12pt]{article}
\usepackage{latexsym,amsmath,amssymb,graphicx}
\renewcommand{\thefootnote}{\fnsymbol{footnote}}

\usepackage{mathrsfs }
\usepackage{amsmath}
\usepackage{amsfonts}

\numberwithin{equation}{section}
	
\def\doubleset#1#2{\bgroup%
\def\doit#1#2{%
\setbox\dblsetbox=\hbox{$\cstyle #1$}%
\raise#2\ht\dblsetbox\copy\dblsetbox%
\hskip-\wd\dblsetbox%
\raise-#2\ht\dblsetbox\box\dblsetbox}%
\mathchoice%
{\def\cstyle{\displaystyle}\doit#1#2}%
{\def\cstyle{\textstyle}\doit#1#2}%
{\def\cstyle{\scriptstyle}\doit#1#2}%
{\def\cstyle{\scriptscriptstyle}\doit#1#2}\egroup}
\def\underarrow#1{\vbox{\ialign{##\crcr$\hfil\displaystyle
 {#1}\hfil$\crcr\noalign{\kern1pt\nointerlineskip}$\longrightarrow$\crcr}}}

\def\IL{\relax{\rm I\kern-.18em L}}
\def\IH{\relax{\rm I\kern-.18em H}}
\def\IR{{\mathbb R}}
\def\IC{\mathbb C}
\def\IB{\relax{\rm I\kern-.18em B}}
\def\ID{\relax{\rm I\kern-.18em D}}
\def\IE{\relax{\rm I\kern-.18em E}}
\def\IF{\relax{\rm I\kern-.18em F}}

\def\IG{\relax\hbox{$\inbar\kern-.3em{\rm G}$}}
\def\IGa{\relax\hbox{${\rm I}\kern-.18em\Gamma$}}
\def\IH{\relax{\rm I\kern-.18em H}}
\def\II{\relax{\rm I\kern-.18em I}}
\def\IK{\relax{\rm I\kern-.18em K}}
\def\IP{\relax{\rm I\kern-.18em P}}
\def\IQ{\relax\hbox{$\inbar\kern-.3em{\rm Q}$}}

\def\CM {{\cal M}}
\def\CN {{\cal N}}

\def\inbar{\,\vrule height1.5ex width.4pt depth0pt}

\newbox\dblsetbox

\newcommand{\dirac}{D\mspace{-13mu}/\mspace{4mu}}

\newbox\dblsetbox

\newlength{\extraspace}
\newlength{\extraspaces}
\newcommand{\be}{\begin{equation}
\addtolength{\abovedisplayskip}{\extraspaces}
\addtolength{\belowdisplayskip}{\extraspaces}
\addtolength{\abovedisplayshortskip}{\extraspace}
\addtolength{\belowdisplayshortskip}{\extraspace}}
\newcommand{\ee}{\end{equation}}

\newcommand{\ba}{\begin{eqnarray}
\addtolength{\abovedisplayskip}{\extraspaces}
\addtolength{\belowdisplayskip}{\extraspaces}
\addtolength{\abovedisplayshortskip}{\extraspace}
\addtolength{\belowdisplayshortskip}{\extraspace}}
\newcommand{\ea}{\end{eqnarray}}

\newcommand{\bd}{\begin{displaymath}
\addtolength{\abovedisplayskip}{\extraspaces}
\addtolength{\belowdisplayskip}{\extraspaces}
\addtolength{\abovedisplayshortskip}{\extraspace}
\addtolength{\belowdisplayshortskip}{\extraspace}}
\newcommand{\ed}{\end{displaymath}}

\newcounter{saveeqn}

\newcommand{\newsection}[1]{
\vspace{12mm} \pagebreak[3] \addtocounter{section}{1}
\setcounter{equation}{0} \setcounter{subsection}{0}
\noindent{\bf \thesection. #1} \nopagebreak
\medskip
\nopagebreak
\addcontentsline{toc}{section}{\thesection. #1}}

\newcommand{\newsubsection}[1]{
\vspace{0.8cm} \pagebreak[3] \addtocounter{subsection}{1}
\setcounter{subsubsection}{0}
\noindent{ \it \thesubsection. #1} \nopagebreak \vspace{2mm}
\nopagebreak
\addcontentsline{toc}{subsection}{\thesubsection. #1}}

\begin{document}
\addtolength{\baselineskip}{1.5mm}

\thispagestyle{empty}

\vbox{} \vspace{1.5cm}

\begin{center}
\centerline{\LARGE{Notes On The ``Ramified'' Seiberg-Witten}}
\bigskip
\centerline{\LARGE{Equations And Invariants}}
\bigskip

\vspace{1.5cm}

{Meng-Chwan~Tan \footnote{e-mail: mengchwan@theory.caltech.edu}}
\\[2mm]
{\it California Institute of Technology, \\
Pasadena, CA 91125, USA} \\ [1mm]
and \\
{\it Department of Physics\\
National University of Singapore \\
Singapore 119260}\\[8mm]
\end{center}

\vspace{1.5 cm}

\centerline{\bf Abstract}\medskip \noindent

In these notes, we carefully analyze the properties of the ``ramified'' Seiberg-Witten equations associated with supersymmetric configurations of the Seiberg-Witten abelian gauge theory with surface operators on an oriented closed four-manifold $X$.  We find that in order to have sensible solutions to these equations, only surface operators with certain parameters and embeddings in $X$, are admissible. In addition,  the corresponding ``ramified'' Seiberg-Witten invariants on $X$ with positive scalar curvature and $b^+_2 > 1$, vanish, while if $X$ has $b^+_2 = 1$, there can be wall-crossings whence the invariants will jump. In general, for each of the finite number of basic classes that corresponds to a moduli space of solutions with zero virtual dimension,  the perturbed ``ramified'' Seiberg-Witten invariants on K\"ahler manifolds will depend --  among  other parameters associated with the surface operator --  on the monopole number $l$ and the holonomy parameter $\alpha$. Nonetheless, the  (perturbed) ``ramified'' and ordinary invariants are found to coincide, albeit up to a sign, in some examples.

\renewcommand{\thefootnote}{\arabic{footnote}}
\setcounter{footnote}{0}

\tableofcontents

\newsection{The ``Ramified'' Seiberg-Witten Equations}

\newsubsection{The Ordinary Seiberg-Witten Equations}

Let us consider the topological Seiberg-Witten (SW) gauge theory, which is a twisted ${\mathcal N}= 2$, $U(1)$ theory on a four-manifold $X$ coupled to a twisted massless hypermultiplet whose lowest component corresponds to a magnetically charged monopole field~\cite{Moore-Witten}. The supersymmetric configurations of the twisted theory are obtained by setting the supersymmetric variations of the fermi fields to zero, and they correspond to the celebrated Seiberg-Witten equations. In order to describe these equations, let us first elaborate on the structure of $X$ and its various associated bundles that are relevant to their description.  

Let $X$ be an oriented, closed four-manifold on which we pick a Riemannian
structure with metric tensor $\bar g$.
$\Lambda^pT^*X$,  or rather $\Lambda^p$,
will denote the bundle of real-valued $p$-forms,
and $\Lambda^{2,\pm}$ will be  the sub-bundle of $\Lambda^2$ consisting
of  self-dual or anti-self-dual forms. If we choose $w_2(X)=0$,
then $X$ is a spin manifold and one can pick
positive and negative spin bundles
$S^+$ and $S^-$, of rank two. Next, let us introduce a  $U(1)$-bundle $L$; the data in the Seiberg-Witten (SW) equations will then be a connection $A$ on $L$ and a monopole field $M$ that is a section of $S^+\otimes L$.  The curvature two-form
of $A$ will be called $F$; its self-dual and anti-self-dual
projections will be called $F_+$ and $F_-$. Since $L$ exists as an ordinary line bundle for $X$ spin, $c_1(L) \in H^2(X, \mathbb Z)$.

If $w_2(X) \neq 0$, that is, if $X$ is not spin, then $S^{\pm}$ do not exist. Nevertheless, a ${\rm Spin}_c$-structure exists on any oriented $X$~\cite{hirz}, and it can be described as a choice of a rank-two complex vector bundle $S^+\otimes L$. In this situation, $L$
does not exist as a line bundle, but $L^{\otimes 2}$ does. (A physical, path integral demonstration of these statements can be found in~\cite{S-duality}, while a similar demonstration for the case with surface operators can be found in~\cite{mine}.)
The data of the SW equations
are now a monopole field $M$ that is a section of $S^+\otimes L$, and a connection on $S^+\otimes L$ with the trace of its curvature form being $2F$. Thus, we see that because the monopole field $M$ is coupled to the gauge field $A$ -- that is, it is a section of $S^+\otimes L$ instead of  $S^+$ -- it is still well-defined on a non-spin manifold. 

At any rate, the SW equations are given by~\cite{monopoles}
\begin{eqnarray}
 {F}^+_{ij}& = & -{i\over 2}\overline M\Gamma_{ij}M   \nonumber \\
   \sum_i\Gamma^iD_iM & = & 0,
  \label{reg SW}
\end{eqnarray}
 where the $\Gamma_i$'s are  Clifford matrices  
(with anticommutators $\{\Gamma_i,\Gamma_j\}=2{\bar g}_{ij}$), such that $\Gamma_{ij}={1\over 2}[\Gamma_i,\Gamma_j]$. In the second equation, $\sum_i\Gamma^iD_i$ is the Dirac operator
$D$ that maps sections of $S^+\otimes L$ to sections of $S^-\otimes L$. Since $S^+$ is
pseudo-real, it will mean that
if $M$ is a section of $S^+\otimes L$, then the complex conjugate $\overline M$
is a section of $S^+\otimes L^{-1}$. Consequently, since the product  $M\otimes \overline M$ lies in $(S^+\otimes L)\otimes (S^+\otimes L^{-1})\cong
\Lambda^0\oplus\Lambda^{2,+}$, and since 
$F_{+}$ also takes values in $\Lambda^{2,+}$, we find that the first equation indeed makes sense.  In terms of positive and negative Weyl-spinor indices $\dot A, A = 1\dots 2$, the SW equations can also be expressed as 
\begin{eqnarray} 
F_{\dot A  \dot B} & = & {i\over 2}\left(M_{ \dot A}\overline M_{ \dot B} +M_{ \dot B}\overline M_{ \dot A} \right) \nonumber \\
D_{A  \dot A} M^{ \dot A} & = & 0.
\end{eqnarray}

\newsubsection{Supersymmetric Surface Operators}
 
Let us now include a supersymmetric surface operator in the theory, which is defined by a field configuration that is a solution to the Seiberg-Witten equations with a singularity along a two-cycle $D$ in $X$. Such a configuration can be represented by  a $U(1)$ gauge field which takes the form 
\be
A = \alpha d\theta + \dots
\label{A}
\ee 
near $D$, where the ellipses refer to the original terms that are regular near $D$,  $\alpha$ takes values in the (real) Lie algebra ${\frak u}(1)$ such that $e^{2\pi \alpha} \in U(1)$, and $\theta$ is the angular variable of the coordinate $z = re^{i \theta}$ of the plane normal to $D$. The resulting field strength is
\be
F = 2\pi \alpha \delta_D + \dots,
\ee  
where $\delta_D$ is a delta two-form that is Poincar\'e dual to $D$. As required, the field strength is singular as one approaches $D$. Moreover, the holonomy in the gauge field is $\textrm{exp}(-2\pi \alpha)$ as one traverses a loop linking $D$. This nontrivial holonomy physically characterizes the surface operator.  Notice that the holonomy is trivial for integer values of $\alpha$;  therefore, $\alpha$ \emph{effectively} takes values in $\mathbb R/ \mathbb Z = S^1 \cong U(1)$. This restriction on $\alpha$ is just the abelian version of the fact that for a non-abelian gauge group $G$, $\alpha$ ought to take values in the maximal torus $\mathbb T \subset G$ rather than its Lie algebra $\frak t$~\cite{Gukov-Witten}. (Note that $\mathbb T = {\frak t} / \Lambda_{\rm cochar}$, where $\Lambda_{\rm cochar}$ is the cocharacter lattice of $G$. In our case, $\mathbb R$ and $\mathbb Z$ play the roles of $\frak t$ and $\Lambda_{\rm cochar}$, respectively.)

The usual physical prescription employed in interpreting operators which introduce a singularity in the gauge field along a two-cycle $D$ in $X$, is to consider in the path integral, connections on the $U(1)$-bundle $L$ (which is singular along $D \subset X$) restricted to $X \backslash D$. In other words, one ought to sum over smooth connections in the path integral if the underlying physical theory is to be finite and therefore well-defined.\footnote{Notice that the above-described surface operator is just a two-dimensional analog of an 't Hooft  loop operator, and in the case that one inserts an 't Hooft loop operator in $X$ -- which introduces a singularity in the gauge field along a loop $\gamma \subset X$ -- one can show that in order to have a well-defined theory, we must sum over all connections of the $U(1)$-bundle (that has a singularity along $\gamma$) over $X \backslash \gamma$  in the path-integral (see $\S$10.3 of~\cite{QFT2}). One does likewise here.} This is \emph{physically equivalent }  (see derivation of eqn.~(2.36) in~\cite{Gukov-Witten}) to considering connections on a $U(1)$-bundle $L'$ which has (smooth) curvature $F' = F - 2\pi \alpha \delta_D$ that extends over all of $X$, where $F$ is the (singular along $D$) curvature of $L$.\footnote{To further justify the arguments in~\cite{Gukov-Witten}, note that the instanton number $\tilde k$ of the bundle $L$ over $X \backslash D$ is (in the mathematical convention)  given by $\tilde k = k +  \alpha l - (\alpha^2 / 2) D \cap D$, where $k$ is the instanton number of the bundle $L$ over $X$ with curvature $F$, and $l = \int_D F / 2\pi$ is the monopole number  ($\it cf$. eqn.~(1.7) of~\cite{KM1} for a $U(1)$-bundle). On the other hand,  the instanton number $k'$ of the bundle $L'$ over $X$ with curvature $F' = F - 2\pi \alpha \delta_D$ is (in the physical convention) given by $k' = - {1\over 8 \pi^2} \int_X F' \wedge F' = k +  \alpha l - (\alpha^2 / 2) D \cap D$. Hence, we find that the expressions for $\tilde k$ and $k'$ coincide, reinforcing the notion that the bundle $L$ restricted to $X \backslash D$ can be equivalently interpreted as the bundle $L'$ defined over all of $X$. Of course, for $F'$ to qualify as a nontrivial field strength, $D$ must be a homology cycle of $X$, so that $\delta_D$ (like $F$) is in an appropriate cohomology class  of $X$.}  In short, in order to introduce a surface operator, one just needs to replace the expression of the field strength that appears in the original Lagrangian (without the surface operator) with $F' = F - 2\pi \alpha \delta_D$, and consider integrating $F'$ over all of $X$ in evaluating the action. Since the positive-definite kinetic terms of the gauge field in the equivalent action are non-singular, the contributions to the path integral will be non-vanishing, as required of a well-defined theory.

\newsubsection{The ``Ramified'' Seiberg-Witten Equations}

The supersymmetric variations of the fields in the presence of a surface operator are the same as those in the ordinary theory without, except that the expression of the (non-singular) field strength is now given by $F'$. As such, the supersymmetric configurations of the theory obtained by setting the variations of the fermi fields to zero can be written as (\ref{reg SW}), albeit in terms of $F'$; in other words, the ``ramified'' SW equations will be given by
\begin{eqnarray}
 ({F - 2\pi \alpha \delta_D})^+_{ij}& = & -{i\over 2}\overline M\Gamma_{ij}M   \nonumber \\
  \sum_i\Gamma^iD_iM & = & 0.
  \label{ram SW}
\end{eqnarray}
 The above equations are consistent with the fact that the monopole field $M$ is  effectively charged under a $U(1)$ gauge field $A'$ with field strength $F'$; that is, $M$ is interpreted as a section of $S^+ \otimes L'$.

In terms of positive and negative Weyl-spinor indices $\dot A, A = 1\dots 2$, the ``ramified'' SW equations can also be expressed as 
\begin{eqnarray} 
(F- 2\pi \alpha \delta_D)_{\dot A  \dot B} & = & {i\over 2}\left(M_{ \dot A}\overline M_{ \dot B} +M_{ \dot B}\overline M_{ \dot A} \right) \nonumber \\
D_{A  \dot A} M^{ \dot A} & = & 0.
\end{eqnarray}   
 
 Last but not least, one can indeed see that since (\ref{reg SW}) holds in the original theory without surface operators, then (\ref{A}) which defines the surface operator, must be a solution to (\ref{ram SW}). In other words, (\ref{A}) defines a supersymmetric surface operator that is compatible with the underlying ${\cal N} =2$ supersymmetry of the SW theory.

 \vspace{-0.5cm}
\newsection{ The ``Ramified'' Seiberg-Witten Invariants}
\vspace{-0.5cm}

\def\cmx{{\CM^{x'}_{\rm sw}}}

\newsubsection{Moduli Space and the ``Ramified'' Seiberg-Witten Invariants}

Let us now describe the moduli space $\CM^{x'}_{\rm sw}$ of solutions to the ``ramified'' SW equations modulo gauge transformations, starting with the dimension. 

In order to compute the (virtual) dimension of the moduli space, first consider (as in~\cite{monopoles}) the following elliptic complex obtained via a linearization of the ``ramified'' SW equations (\ref{ram SW})
\be
0\to \Lambda^0\underarrow{s}\Lambda^1
\oplus (S^+\otimes L')\underarrow{t}\Lambda^{2,+}
\oplus (S^-\otimes L') \to 0,
\ee
where $t$ is the linearization of the ``ramified'' SW equations, and $s$ is the map from zero forms to deformations in $A'$ (the connection on $L'$) and $M$ induced by the infinitesimal action of the $U(1)$ gauge group. 
Next, define the operator $T = s^* \oplus t$ (where $s^*$ is the adjoint of $s$), which can be described as the map
\be
T:\Lambda^1\oplus(S^+\otimes L')\to \Lambda^0\oplus \Lambda^{2,+}
\oplus (S^-\otimes L').
\ee
Then, the (virtual) dimension of the moduli space is given by the index of $T$.
By dropping terms in $T$ of order zero,
$T$ can be deformed to the direct sum of the operator $d+d^*$ (projected onto self-dual two-forms)
from $\Lambda^1$ to $\Lambda^0\oplus \Lambda^{2,+}$ and the (twisted) Dirac
operator $\dirac$ from $S^+\otimes L'$ to $S^-\otimes L'$, where $d^*: \Lambda^1 \to \Lambda^0$ is the adjoint of $d$.
The index of $T$ is
the index of $d+d^*$ plus twice the index of the $\dirac$. The $d+d^*$ operator is independent of $A'$  and $M$ because the group is abelian, and its index is given by $- (\chi + \sigma) / 2$. On the other hand, the operator $\dirac$ depends on $L'$ and hence $A'$; the expression for twice its index  is $- \sigma /4 + c_1(L')^2$. Consequently, the index of $T$, which gives the (virtual) dimension of $\cmx$, is
\be
d^{x'}_{\rm sw} = - {{2 \chi + 3 \sigma} \over 4} + c_1(L')^2.
\label{dimension}
\ee

\def\dsw{{d^{x'}_{\rm sw}}}

For there to be solutions to the ``ramified'' SW equations, we must have $\dsw \geq 0$. Moreover, since $\dsw$ is the dimension of $\cmx$, it must always be an integer. In the special case that $\dsw=0$, that is, when $x' = -c_1(L'^2) = -2c_1(L')$   obeys 
\be
x'^2 = 2 \chi + 3 \sigma,
\label{x2}
\ee  
the moduli space generically consists of a finite set of points $P_{i,x'}$, $i=1\dots t_{x'}$.
With each such point, one can associate a number $\epsilon_{i,x'}$  given by the sign of the determinant of $T$. The ``ramified'' SW invariant  corresponding to such a particular choice of $x'$ is then 
\be
SW({x'}) =\sum_i \epsilon_{i,x'}.
\label{SW inv for d=0}
\ee
As in the ordinary case, let us call such an $x'$ a basic class. We will show later that there are only a finite number of basic classes which correspond to $SW({x'}) \neq 0$, and that $SW ({x'})$ is a topological invariant if $b^+_2 (X) \geq 2$.  

Note that $\dsw=0$ if and only if the index of the Dirac operator
is 
\be
\Delta={\chi+ \sigma\over 4}.
\ee
Since $\Delta$ must be an integer, it implies that \emph{only manifolds with integral values of ${(\chi+ \sigma) / 4}$ have nontrivial $SW({x'})$.} Also, note that 
we have 
\be
-\alpha l + {\alpha^2 \over 2} D \cap D =   {\sigma \over 8} + \Delta + k,
\label{condition1}
\ee
where $k = -{1\over 8 \pi^2} \int_X F \wedge F$ and $l = \int_D F / 2\pi$ are integers. Therefore, we find that  the integrality of $\Delta$ also implies that
\be
-\alpha l + {\alpha^2 \over 2} D \cap D =   {\sigma \over 8} \  \mathrm{mod} \ 1.
\label{condition}
\ee
In other words, \emph{surface operators that lead to nontrivial  $SW({x'})$ will have parameters $\alpha$, $l$  and self-intersection numbers $D \cap D$ that obey (\ref{condition1}) and hence, (\ref{condition}).}

Notice that if we replace $L$ by ${L'}^{-1}$ (and therefore $A'$ by $-A'$), and $M$ by $\overline M$,  the ``ramified'' SW equations are invariant. Nevertheless, the sign of the determinant of $T$ will be multiplied by $(-1)^{\Delta}$ ($\it cf$.~\cite{monopoles}). Hence, we find that 
\be
SW(- x') = (-1)^{\Delta} SW(x').
\label{swop}
\ee            

In the general case where $\dsw > 0$, $x'$ will no longer be given by (\ref{x2}), and the ``ramified'' SW invariants will be given by ($cf$.~\cite{Marcos}) 
\be
SW_{x'} (\beta_{i_1} \wedge \dots \wedge \beta_{i_r})  = \int_{{\cal M}^{x'}_{\rm sw}} \nu_{i_1} \wedge \dots \wedge \nu_{i_r} \wedge a^{{1\over 2} (\dsw - r)}_d, 
\label{SW invariants}
\ee
where $\dsw - r$ must be  even and $\it{positive}$, $a_d$ is the vacuum expectation value of the complex scalar $\varphi_d$ in the  $\CN =2$ ``magnetic'' $U(1)$ vector multiplet, and
\be
\nu_i \sim \int_{\delta_i} \psi_d,
\ee
where $\psi_d$ is a (spacetime) one-form fermi field that is also in the  $\CN =2$ ``magnetic'' $U(1)$ vector multiplet. Also, $\delta_1, \dots, \delta_r \in H_1(X, \mathbb Z)$ with duals $\beta_1 \dots \beta_r \in H^1(X, \mathbb Z)$.  Alternatively, the cokernel of $T$ will, in this case, be a vector bundle $V$  over $\cmx$, and its Euler class integrated over $\cmx$ will give us $SW_{x'} (\beta_{i_1} \wedge \dots \wedge \beta_{i_r}) $. 

In the case of $\dsw=0$ (where the invariants can only be defined as (\ref{SW inv for d=0})), there will potentially be a difference between the ordinary and ``ramified'' SW invariants  if the sign of the determinant of $T$ involves $F'$; it is only through $F'$ that the invariants can inherit a dependence on $\alpha$ and $l$, which, then, distinguishes them from the ordinary invariants. On the other hand, for $\dsw > 0$, the expression of the invariants in (\ref{SW invariants}) already manifestly depends on $\alpha$ and $l$ through $\dsw$, since
\be
\dsw  = \left ( {\alpha^2} D\cap D -2 \alpha l \right)  - 2\left ({2 \chi + 3 \sigma \over 8}  + k \right) . 
\ee

\bigskip\noindent{\it Orientation of the Moduli Space}

In order for the above discussion to be technically consistent, one still needs to pick an orientation on $\cmx$. Equivalently,  one needs to pick an orientation of the cohomology groups ${\cal H}^0 = H^0(X, \IR)$, ${\cal H}^1 = \textrm{Ker} \dirac \oplus H^1(X, \IR)$ and ${\cal H}^2 = \textrm{Coker} \dirac \oplus H^{2,+}(X, \IR)$  of the (deformed) elliptic complex~\cite{Scorpan}. 

The line $H^0(X, \IR)$ has a canonical orientation, given by the class of the constant function $x \mapsto 1$, where $x$ is a point in $X$. That leaves us with the orientations of ${\cal H}^1$ and ${\cal H}^2$. Since $\dirac$ is an elliptic, $\IC$-linear operator, both $\textrm{Ker} \dirac$ and $\textrm{Coker} \dirac$ are finite-dimensional complex vector spaces; thus, they have natural orientations of their own. It is therefore clear that in order to pick an orientation on $\cmx$, one just needs to pick an orientation of  $H^1(X, \IR) \oplus H^{2,+}(X, \IR)$, as in the ordinary case.

\newsubsection{Wall-Crossing Phenomenon}

Generically, $SW(x')$, which `counts' the number of solutions to the ``ramified'' SW equations (weighted by a sign), is a topological invariant. Nevertheless, under certain special conditions, it can jump as one crosses a ``wall'' while moving in the space of metrics on $X$. This subsection is devoted to explaining this in detail. 

\medskip\noindent{\it A Relevant Digression}

Before we proceed further, let us discuss something that was implicit in our discussions hitherto: that $L'$ exists as a (complex) line bundle over $X$ which  is spin, so $c_1(L') \in H^2(X, \mathbb Z)$.  In other words, $\int_U (F' / 2\pi) \in \mathbb Z$ for $\it any$ integral homology 2-cycle $U \subset X$ (assuming, for simplicity, that the homology of $X$ is torsion-free), and therefore, $\alpha  (U \cap D)   \in   \mathbb Z$ in any physically sensible solution. Note that this condition is consistent with the fact~\cite{Gukov-Witten} that one can invoke a twisted $U(1)$-gauge transformation in the physical theory -- that leaves the holonomy ${\rm exp}(-2\pi \alpha)$ and thus the \emph{effective} ``ramification'', invariant -- which shifts $\alpha \to \alpha + u$ for some non-integer $u$, such that non-trivial values of $\alpha$ can, in this context, be regarded as integers; in particular, this also means that $\alpha  (D \cap D)   \in   \mathbb Z$ -- a condition which underlies the integrality of the monopole number $l$.

\medskip\noindent{\it Wall-Crossing Phenomenon}

Now, since $M$ is charged under the gauge field, it is acted upon by gauge transformations.  Consequently, any solution to the ``ramified'' SW equations  with $M=0$  represents  a fixed, singular point  in $\cmx$ -- the space of all solutions modulo gauge transformations. As such, it would -- as in the case with Donaldson theory~\cite{Donaldson} -- result in a jump in $SW(x')$ as one crosses a ``wall'' while moving in the space of metrics on $X$.  In other words, $SW(x')$ will fail to be a topological invariant if there is a nontrivial solution to $F'_+ = 0$ -- the ``ramified'' abelian instanton.  Let us ascertain when such a nontrivial ``ramified'' abelian instanton exists.  

As   $(F' / 2 \pi) \in H^2(X, \mathbb Z)$, the condition $F'_+ = 0$ implies that $F'/ 2\pi$ lies at the intersection of the integral lattice in $H^2(X, \IR)$ and its anti-self-dual subspace $H^{2,-}(X,\IR)$. As long as $b^+_2 \geq 1$, so that the self-dual part of $H^2(X, \IR)$ is non-empty, the intersection in question just consists of the zero-vector. Hence, for a generic metric on $X$, there are no ``ramified'' abelian instantons.  

Nonetheless, for $SW(x')$  to qualify as a genuine topological invariant, it will mean that along any path that connects two generic metrics on $X$, there cannot be a ``ramified'' abelian instanton. This can fail for $b^+_2 =1$, as in this case, the dimension of the self-dual subspace of $H^2(X, \IR)$ is one, and in a generic one-parameter family of metrics on $X$, one may meet a special metric whereby the above-mentioned intersection is non-zero, such that $SW(x')$ can then jump.
Let us analyze how this happens (following~\cite{monopoles}), assuming for simplicity that $b_1=0$ and
$b_2 = b_2^+=1$.

Even though the equation $F'_+(A') = 0$ has no solutions for a generic metric on $X$,  we would like to investigate what happens when we are close to a special metric where there is one.  To this end, let us parameterize the family of metrics on $X$ by $\epsilon$, such that at  a ``wall'' where $\epsilon =0$, we have a solution $A'_0$ to $F'_+(A') = 0$. In other words,  at $\epsilon =0$, there is a solution $A' = A'_0$ and $M=0$ to the ``ramified'' SW equations.  What we would like to study, is the solution to the equations for a nearby metric corresponding to a small, non-zero $\epsilon$. At this point, note that one has $\dsw = 0$ precisely if the index $\Delta$ of the Dirac
equation is 1.  Therefore, there is generically a single (non-zero) solution $M_0$
of the Dirac equation $\sum_i \Gamma^iD_iM=0$ at $\epsilon \neq 0$. As such, near $\epsilon =0$, we can write the solution to the equations at $\epsilon \neq 0$ as $A' = A'_0 + \epsilon \delta A'$ and $M = m M_0$, where $m$ is a complex number, and   
\be
F'^+_{ij} (A_0) + \epsilon (d \delta A')^+_{ij}    + {i\over 2} m \overline m\overline M_0\Gamma_{ij}M_0   = 0.
\label{eqns}
\ee
 Since $b_1 =0$, it will mean that one can always make a choice of $\delta A'$ which is $d$-exact. Moreover, since $F'_+ (A_0)= 0$ at $\epsilon =0$, it will mean that $F'_+ (A_0)$ is proportional to  $\epsilon$. Hence, (\ref{eqns}) can also be written as
\be
c \epsilon =  |m|^2,
\label{wall-cross}
\ee
where $c$ is some constant.     

Clearly, since $|m|^2$ is always positive, it will mean from (\ref{wall-cross}) that  \emph{for $b_1 =0$ and $b^+_2 =1$, the number of solutions $SW(x')$ to the ``ramified'' SW equations will jump by $+1$ or $-1$ as one crosses a ``wall'' in going from $\epsilon < 0$ to $\epsilon >0$,  depending on the sign of $c$.}  

Last but not least, at the ``wall'' where $F'_+ =0$ is supposed to be a solution of the ``ramified'' SW equations, we must have $\dsw \geq 0$. Since $F'_+ = 0$ means that $\ast x' = -x'$, (where $\ast$ is the Hodge-dual operator), we also have ${x'}^2 <0$. Then, for $b_1 =0$ and $b^+_2 =1$, we find that 
\be
\label{wallcrossing condition}
 (2 +  {\sigma \over 4}) \leq \left( {\alpha^2} D\cap D -2 \alpha l  -2k \right) < 0.
\ee
Hence, \emph{for ``ramified'' SW wall-crossing to occur at a particular metric on $X$, the corresponding values of $\alpha$, $D \cap D$, $l$ and $k$ must obey the inequality above.}  In particular, \emph{the location of the ``wall'' will now depend  on  the parameter $\alpha$.}

\newsubsection{Properties of the ``Ramified'' Seiberg-Witten Invariants and their Basic Classes}

\medskip\noindent{\it Properties of the ``Ramified'' SW invariants}

Let us now explore the properties of the ``ramified'' SW invariants through some vanishing theorems.  As a start, let us first schematically write the ``ramified'' SW equations as $F'_+ = \phi(M)$ and ${\dirac} M = 0$. Next,  note that from the coupled Lichnerowicz formula~\cite{Morgan}, we have
\be
{\dirac}^{\dagger} {\dirac} M = D^{\dagger} D M + {1 \over 4}R  \cdot  M + {1\over 2}  (F'_+ \star M),
\label{coupled Lichnerowicz}
\ee
where $D_i = \partial_i + w_i  + A'_i$ is the covariant derivative with respect to the spin and gauge connections $w_i$ and $A'_i$, $R$ is the scalar curvature of $X$, and $\star$ denotes the Clifford multiplication.  Note that the adjoint of the (twisted) Dirac operator ${\dirac}^{\dagger}$ acts on sections of $S^- \otimes L'$, that is, $\overline M$. Also, note that since $M$ (unlike $F$) is a non-singular (monopole) field, there will not be any surface contributions along $D$ in the integral over $X$ of its (covariant) derivative. Altogether, this means that we can write
\be
 I = \int_X d^4x \sqrt {\bar g} \ |\dirac M|^2 = \int_X d^4x \sqrt {\bar g} \left( |D M|^2 + {1\over 4} R \cdot |M|^2 + {1 \over 4} |M|^4 \right),
 \label{I}
\ee 
where we have used the fact that $(\phi (M) \star M ) = {1\over 2} |M|^2 \cdot M$. In the above, $|\Omega|^2 = \langle \Omega  , \Omega \rangle$, where $\langle , \rangle$ is the inner product relevant to the field $\Omega$.  

Since we have, at the outset, assumed that $F'_+ = \phi(M)$ in writing $I$, the vanishing of $\dirac M$ and hence $I$, would then correspond to a solution of  the ``ramified'' SW equations. In particular, for $R > 0$, a solution would be given by $M =0$ and $F'_+ =0$.  
For $b^+_2 \geq 2$, we have already seen that there are no nontrivial solutions to $F'_+ =0$.  In other words, \emph{for manifolds $X$ with positive scalar curvature and $b^+_2 \geq 2$, we have $SW(x') = 0$.} This statement holds for all values of $\alpha$; in particular, for $\alpha \in \mathbb Z$, that is, when  ``ramification'' is effectively absent, our observation just reduces to that of the \emph{ordinary} SW invariants established in~\cite{monopoles}.   On the other hand, for $b^+_2 =1$, there can be nontrivial solutions to $F'_+ =0$ at a ``wall''.  Hence, \emph{if the condition (\ref{wallcrossing condition}) is satisfied, $SW(x') \neq 0$ for some metric on $X$ with $b_1 =0$, $b^+_2 = 1$ and positive scalar curvature}.

What can we say when $X$ has $R=0$? Let us consider  a perturbation of the ``ramified'' SW equations, which now read $F'_+ = \phi(M) - \rho_+$ and $\dirac M =0$, where $\rho_+$ is a perturbation self-dual harmonic two-form like $F'_+$;  the data of the perturbed ``ramified'' SW equations can then be identified with a point in the space of pairs $(\bar g, \rho_+)$ on $X$. Next, perturb the metric $\bar g$ to another metric ${\bar g}'$.  Altogether, $I$ will be replaced by
\be
 {I}' = \int_X d^4x \sqrt {{ \bar g}'} \ |\dirac M|^2 = \int_X d^4x \sqrt {{\bar g}'} \left( |D M|^2 + {1\over 4} R \cdot |M|^2 + {\sqrt 2 \over 4} |M|^2 \cdot |\phi(M)- \rho_+| \right).
 \label{perturbed}
\ee 
If $R> 0$, we find from $I'$ that a possible solution would be $M=0$ and $F'_+ = - \rho_+$. However, a generic choice of $\rho_+$ 
cannot coincide with $F'_+$ itself; as such, there are generically no solutions when $R> 0$. Nevertheless, for $R=0$, notice that there is a solution given by $DM = 0$ and $\phi(M) = \rho_+$. In other words, this solution is characterized by a ``ramified'' abelian instanton $F'_+ = 0$, and a covariantly-constant $M$ and thus $\rho_+$. For all $\rho_+ \in H^{2, +}(X, \mathbb Z)$ to be covariantly-constant,  $X$ must either be K\"ahler with $b^+_2 = 1$ (such that one can actually meet a solution of $F'_+ = 0$ for some $({\bar g}', \rho_+)$)  or  is hyper-K\"ahler. Therefore, \emph{we find that the (perturbed) ``ramified'' SW invariants are non-zero on scalar-flat K\"ahler manifolds with $b^+_2 =1$, and on hyper-K\"ahler manifolds.} Indeed,  as shown in~\cite{le brun}, one can find scalar-flat K\"ahler metrics on non-minimal ruled surfaces with $b^+_2 =1$. Moreover, hyper-K\"ahler manifolds are necessarily scalar-flat. 

\bigskip\noindent{\it The ``Ramified'' SW Invariants for Connected Sums}

\emph{Suppose that $X$ is the connected sum of two manifolds $N'$ and $N''$ -- that is, $X = N' \# N''$, with both $b^+_2(N') \geq 1$ and $b^+_2(N'') \geq 1$.}  We wish to ascertain $SW(x')$ on such an $X$. Since $SW(x')$ is supposed to be a topological invariant, the result will be unaffected if we stretch the cylinder which connects $N'$ and $N''$ in $X$. The topology of the elongated cylinder is given by $S^3 \times [0,1]$, and can thus be arranged to have $R > 0$, since $S^3$ has positive scalar curvature.  Based on our earlier analysis, there are no solutions to the ``ramified'' SW equations along the cylinder. This means that solutions to the ``ramified'' SW equations on $X$ must come only from solutions on $N'$ and  solutions from $N''$. This means that the moduli space of solutions on $X$ will be given by
\be
{\cal M}_{X} = {\cal M}_{N'} \times {\cal M}_{N''}. 
\label{moduli space}
\ee
Note that any Chern class of the line bundle $L'$ over $X$ is given by the Chern class of $L'$ over $N'$ plus the Chern class of $L'$ over $N''$. Likewise for the signature $\sigma(X)$. Note also that $\chi(N' \# N'') = \chi(N') + \chi(N'') -2$. Altogether, this means from (\ref{dimension}) that 
\be
\textrm{dim} {\cal M}_X = \textrm{dim} {\cal M}_{N'} + \textrm{dim} {\cal M}_{N''}  + 1.
\ee
Since we are interested in the case where $\textrm{dim} {\cal M}_X = 0$, it must mean that either $\textrm{dim} {\cal M}_{N'}$ or $\textrm{dim} {\cal M}_{N''}$ has (virtual) dimension $-1$, that is, it is empty.\footnote{Recall that because $b^+_2 \geq 1$ for either $N'$ or $N''$, it will mean that for a generic metric, there are no ``ramified'' abelian instantons $F'_+ =0$ and hence, no solutions to $M=0$. Consequently, ${\cal M}_{N'}$ and ${\cal M}_{N''}$ are both smooth manifolds with non-negative (virtual) dimension. As such, they will be empty if found to have negative (virtual) dimension.}  In turn, we find from (\ref{moduli space}) that ${\cal M}_X$ is also empty. \emph{Therefore, $SW(x') =0$ on $X$.}

\bigskip\noindent{\it About the Basic Classes $x'$}  

We shall now demonstrate that there are only a finite number of choices of $x'$ that give rise to non-zero $SW(x')$. To begin, first note that for $SW(x')$ to be non-zero, $I$ in (\ref{I}) must be zero, that is,
\be
 \int_X d^4x \sqrt {\bar g} \left( |D M|^2 + {1\over 4} R \cdot |M|^2 + {1 \over 4} |M|^4\right) = 0.
\ee 
This implies that
\be
{1 \over 4} \int_X  |M|^4 \leq \int_X  \left(|D M|^2  + {1 \over 4} |M|^4\right) =  {1\over 4} \int_X (-R) \cdot |M|^2   \leq  {1\over 4}\left( \int_X {R^2}\right)^{1/2}\left( \int_X |M|^4\right)^{1/2},
\ee
where we used the Cauchy-Schwarz inequality in the last step. Since $|F'_+| \sim |M|^2$, we then find that
\be
\left(\int_X  |F'_+|^2\right)^{1/2} \leq    \left( \int_X \zeta {R^2}\right)^{1/2}
\ee
for some constant $\zeta$.  Therefore, 
\be
\int_X c_1(L')^2  = {1\over 4 \pi^2} \left ( \int_X |F'_+|^2 - \int_X |F'^-|^2 \right) \leq {1\over 4 \pi^2} \int_X |F'_+|^2 \leq  \left( \int_X \zeta' {R^2}\right)
\ee
for some constant $\zeta'$. Finally, since  $d \geq 0$ for $SW(x') \neq 0$, we find from (\ref{dimension}) that
\be
{{2 \chi + 3 \sigma} \over 4} \leq \int_X c_1(L')^2  \leq  \left( \int_X \zeta' {R^2}\right).
\label{range}
\ee
Thus, we see that \emph{there are only a finite number of choices of basic classes $x' = -2c_1(L')$ that result in $SW(x') \neq 0$; namely, those that satisfy (\ref{range}).} 

\vspace{-0.0cm}
\newsection{The ``Ramified'' Seiberg-Witten Invariants on K\"ahler Manifolds}

We shall now specialize to the case where $X$ is K\"ahler, and ascertain the (perturbed) ``ramified'' SW invariants. To this end, we shall adopt a strategy similar to that employed in~\cite{monopoles}.

\vspace{-0.0cm}
\newsubsection{The Moduli Space of The  ``Ramified'' Seiberg-Witten Equations On K\"ahler Manifolds}

\medskip\noindent{\it Constraint on the Embedding $D$ in K\"ahler Manifolds}

If $X$ is K\"ahler and spin (as assumed at the outset), then $M = S^+\otimes L'$ has a decomposition
$S^+\otimes L'\cong (K^{1/2}\otimes L')\oplus (K^{-1/2}\otimes L')$,
where $K$ is the canonical bundle and $K^{1/2}$ is its square root.
(If $X$ is K\"ahler but $\it not$ spin,  $S^+\otimes L'$ can still be decomposed as stated. However, $K^{1/2}$ and $L'$ can no longer exist
separately, and $K^{1/2}\otimes L'$ must be characterized
as a square root of the line bundle $K\otimes{ L'}^2$.) 

Let us denote the components of $M$ in $K^{1/2}\otimes L'$ and
in $K^{-1/2}\otimes L'$ as $\xi$ and $-i\bar \beta$, respectively. The equation $F'_+= \phi(M)$ can now be decomposed as
\begin{eqnarray}
\label{kahler equations}
F'^{2,0} & = & \xi \beta \nonumber \\
{F}_\omega^{'1,1} & = & -{\omega\over 2}  \left(|\xi|^2-|\beta|^2\right) \\
F'^{0,2} & = & \bar\xi \bar\beta.\nonumber
 \end{eqnarray}
Here, $\omega$ is the K\"ahler form and ${F}_\omega^{'1,1}$ is the $(1,1)$
part of $F'_+$. Consequently, (\ref{I}) can be written as
\be
 I_{\textrm {k\"ahler}} = \int_X d^4x \sqrt {\bar g} \ |\dirac M|^2 = \int_X d^4x \sqrt {\bar g} \left( |D \xi|^2 + |D \beta|^2   + {1\over 4} R \cdot (|\xi|^2 + |\beta|^2) + {1 \over 4} (|\xi|^2 + |\beta|^2)^2 \right).
 \label{I_kahler}
\ee 
As before, solutions to the ``ramified'' SW equations would be such that the right-hand-side of (\ref{I_kahler}) vanishes. Notice that for non-zero $\xi$ and/or $\beta$, there can only be solutions to the equations (for non-flat $F')$ if $R < 0$. This agrees with our earlier observation that $SW(x')$ vanishes on manifolds with positive scalar curvature. 

Consider the map 
\begin{eqnarray}
A' & \to & A' \nonumber \\
\xi & \to & \xi \\
\beta & \to & - \beta. \nonumber
\end{eqnarray} 
Notice that the above map is a symmetry of the right-hand-side of (\ref{I_kahler}). This means that for some value of $\xi$ and $\beta$ that (\ref{I_kahler}) is zero, that is, for some solution of the ``ramified'' SW equations, the above map gives another zero of (\ref{I_kahler}) and hence, another solution to the ``ramfied'' SW equations. Notice, that the two solutions related by the map can satisfy (\ref{kahler equations}) only if $\xi \beta$ and $\bar \xi \bar \beta$ are zero, that is, 
\be
F'^{2,0} =  F'^{0,2} = 0.
\ee
This means that the connection $A'$  defines a holomorphic structure on $L'$. The basic classes $x' = -2 c_1(L')$ are therefore of type $(1,1)$ for any K\"ahler structure on $X$. 

This constraints the number of admissible embeddings $D$ of the surface operator in $X$: we find that  \emph{there can only be a total of $h^{1,1} = b_2 - 2 h^{2,0}$ instead of $b_2$  choices of $D$ among the linearly-independent two-cycles in (the torsion-free part of) $H_2(X, \mathbb Z)$ if it is K\"ahler, where $h^{p,q} =  \textrm{dim} [H^{p,q}(X, \IR)]$.}

\bigskip\noindent{\it The Moduli Space $\cmx$}

If $\xi \beta$ (or $\bar\xi \bar\beta$)  is to vanish whilst ${F}_\omega^{'1,1}  \neq 0$, it would mean that either $\xi =0 $ and $\beta \neq 0$, or $\xi \neq 0 $ and $\beta = 0$. Consequently, the second of the ``ramified'' SW equations will read $\bar\partial_{A'} \beta = 0$ or $\bar\partial_{A'} \xi = 0$, respectively, where $\bar \partial_{A'}$ is the $\bar \partial$-operator on $L'$ -- that is, $\xi$ and $\beta$ are holomorphic sections of the appropriate bundles when they do not vanish. 

Let us now consider the case  where $\beta =0$. (The discussion involving $\xi = 0$ is more or less identical; one just has to replace $L'$ with ${L'}^{-1}$ throughout our proceeding discussion).  Then the space of connections $A'$ and sections $\xi$ of the bundle $K^{1/2} \otimes L$ span a symplectic manifold, with symplectic structure defined by 
\begin{eqnarray}
\langle\delta_1A',\delta_2A'\rangle & = & \int_X\omega   \wedge \delta_1A'\wedge\delta_2 A'  \nonumber \\
 \langle \delta_1\xi,\delta_2\xi \rangle & = & -i\int_X\omega\wedge\omega
\left(\delta_1\overline \xi\delta_2\xi-\delta_2\bar \xi \delta_1\xi\right).
\end{eqnarray}
On this symplectic manifold acts the group of $U(1)$ gauge transformations.
The moment map $\mu$ for this action is the quantity that appears in the
$(1,1)$ equation of (\ref{kahler equations}), that is
\be
\mu\omega= F_{\omega}^{'1,1}+ \omega |\xi|^2.
\ee
Consequently, the moduli space $\cmx$ can be obtained by setting
to zero the above moment map and dividing by the group of $U(1)$ gauge
transformations.  This should be
equivalent to dividing by the complexification of the group of gauge
transformations. In the case at hand, the complexification of the
group of gauge transformations acts by $\xi\to \lambda \xi$,
$\bar\partial_{A'}\to \lambda\bar\partial_{A'}\lambda^{-1}$, where $\lambda$ is a map from
$X$ to ${\bf C}^*$.

 Conjugation by $\lambda$ has the effect of identifying any two $A'$'s that
define the same complex structure on $L'$.  Hence, $\cmx$ is the moduli space of
pairs consisting of a complex structure on $L'$, and a non-zero
holomorphic section $\xi$ of $K^{1/2}\otimes L'$ that is defined up to scaling. Furthermore, if $X$ has $b_1=0$, 
then the complex structure on $L'$ (if it exists) is unique.
$\cmx$ will then be a complex projective space, ${\bf P}H^0(X,K^{1/2}\otimes L')$. Nevertheless, since $\xi$ is holomorphic, by the maximum modulus principle~\cite{J.Moore}, $\xi$ is constant. Consequently,  \emph{$\cmx$ consists only of a single point for the canonical basic class $x' = x'_c = c_1(K)$.} 

\newsubsection{The ``Ramified'' Seiberg-Witten Invariants on K\"ahler Manifolds}

\medskip\noindent{\it The Basic Classes $x'$}

Recall from our analysis at the end of the previous section that there are only a finite number of $x'$'s. Generalizing the analysis in $\S$4 of~\cite{monopoles} to the ``ramified'' case, (which simply involves replacing the ordinary  $U(1)$-bundle in~\cite{monopoles} with $L'$), we learn that the $x'$'s  correspond (possibly in a one-to-many fashion) to the independent global sections of the canonical line bundle $K$ of $(2,0)$-forms on $X$. In other words, \emph{the total number of independent basic classes $x'$ will be given by $\textrm{dim} [H^0(X, K)]$, at  most. }  

\def\cmxc{{\CM^{x'_c}_{\rm sw}}}

\bigskip\noindent{\it  The (Perturbed) ``Ramified'' SW Invariants}

Since $\cmxc$ consists only of a single point when $b_1=0$, it will mean from (\ref{SW inv for d=0}) that the ``ramified'' SW invariant for the canonical basic class $x'_c$ is just the sign of the determinant of $T$.  For K\"ahler manifolds $X$ with $b^+_2(X) > 1$, the expression for the sign of the determinant of $T$ is the same as in the ordinary case -- that is, the\emph{ perturbed }SW invariant ${\widetilde {SW}}(x'_c) = (-1)^w$, where $w = \textrm{dim}_{\mathbb C} [\textrm{Ker} T]$. In order to see this,  one just needs to repeat the arguments laid out at the end of $\S$4 of~\cite{monopoles} after replacing the ordinary $U(1)$-bundle therein with $L'$.\footnote{To arrive at this result, we have also made use of the fact that the moduli space of the perturbed SW equations for the basic class $x'_c$  -- like $\cmxc$ -- consists of a single point only.}  Since the relevant computation is based on a standard treatment which involves mildly deforming the determinant of $T$ to the Ray-Singer-Quillen determinant, we shall,  in favor of brevity, not repeat it here. 

At any rate, it should be emphasized that  ${\widetilde {SW}}(x'_c)$ may depend on the additional parameters $\alpha$, $\l$ and $D \cap D$: notice that $w = \textrm{dim}_{\IC}[\textrm{Ker} T]$ depends on the index of $T$, which, in turn, is expressed in terms of these parameters. Nevertheless, since $w$ is necessarily an integer,  \emph{the  ${\widetilde {SW}}(x'_c)$'s for (hyper)-K\"ahler manifolds with $b_1=0$ and $b^+_2  > 1$, are given by $\pm 1$.}

\bigskip\noindent{\it Explicit Dependence of  ${\widetilde {SW}}(x')$ on $\alpha$, $l$ and $D \cap D$}

Ascertaining the explicit dependence of $w$ and hence of the ${\widetilde {SW}}(x')$'s on the surface operator parameters $\alpha$, $l$ and $D \cap D$, is in general difficult, as one can usually compute only the index of $T$ and not its (co)kernel alone. Nevertheless, in certain special cases, the dependence of the ${\widetilde {SW}}(x')$'s on the above surface operator parameters can be made manifest.  

For example, for ruled surfaces given by an $S^2$-bundle over a Riemann surface $\Sigma_g$ of genus $g$, the scalar curvature $R$ is positive (for a metric where $S^2$ is small) and  $b_1 = 2g$. It also has $b^+_2 =1$, and  thus, one can find a solution to $F_+ = -\rho_+ =0$ for some special pair $(\bar g, \rho_+)_s$; this implies that there can be wall-crossings, since $M=0$. The wall-crossing formula of the ordinary perturbed SW equations has been determined in~\cite{Li}, and since the ordinary perturbed SW invariants vanish in some chamber in the space of pairs $(\bar g, \rho_+)$  when $R>0$, the wall-crossing formula in~\cite{Li} actually gives us the exact non-zero expression of the ordinary perturbed SW invariants for some other pair $({\bar g}', \rho'_+)$ in an adjacent chamber. As the ``ramified'' SW equations are just the ordinary SW equations with the ordinary $U(1)$-bundle replaced by $L'$, and since the perturbed ``ramified'' SW invariants also vanish in some chamber in the space of pairs $(\bar g, \rho_+)$  when $R > 0$ (as argued in $\S$2.3), their non-zero values for some other pair $(\bar g', \rho'_+)$ in an adjacent chamber will be given by ($\it cf$.~\cite{Li})   
\be
{\widetilde {SW}}(x')   = \pm \left(\int_{{\bf S}^2} {F \over 2\pi} -   \alpha  ({\bf S}^2 \cap D) \right)^g.
 \label{jump ruled}
\ee

The dependence of ${\widetilde {SW}}(x')$ on the parameter $\alpha$ is manifest; however, the dependence on $l$ and $D \cap D$ is still implicit. Nevertheless, as we shall see when we consider an explicit example below, all parameters can be made manifest in the final expression of ${\widetilde {SW}}(x') $. Also, ${\widetilde {SW}}(x') $ is not necessarily equal to $\pm 1$ unless $g = 0$ -- this is consistent with our earlier analysis, where we showed that for $b_1 =0$ and $b^+_2 =1$, the ``ramified'' invariants will jump by $\pm 1$.\footnote{Actually, we showed this to be true of the unperturbed invariants at the end of $\S$2.2. Nevertheless,  one can generalize the analysis to include the perturbation two-form $\rho_+$, whereby $\rho_+$ necessarily vanishes at $\epsilon =0$, as this is where the intersection of the integral  lattice in $H^2(X, \mathbb R)$ with its anti-self-dual subspace $H^{2,-}(X, \mathbb R)$ is non-zero. In doing so, one will obtain a similar conclusion -- that the perturbed invariants will jump by $\pm 1$ in crossing a ``wall'', if $b_1 =0, b^+_2 =1$.}

\newsubsection{Some Explicit Examples} 

\medskip\noindent{\it The Invariants on $K3$}

Let us consider  a closed hyper-K\"ahler manifold $X$ such as $K3$ which has $b_1=0$, $b^{\pm}_2(X) > 1$ (and $R=0$). Then $\chi(X) = 24$ and $\sigma(X) = -16$. Consequently, ${x'}^2 = 2 \chi + 3 \sigma = 0$; that is, there is only one basic class for $K3$ -- the trivial one $x'_c = c_1(K)$. This is indeed consistent with the fact that $\textrm{dim}[H^0(K3, K)] =1$, where $K$ is trivial. From our above discussion, we have
\be
{\widetilde {SW}}_{K3} (0) = \pm 1.
\ee
Because  $b^+_2 (X) =3$, there are no wall-crossings, and the above result holds for all metrics on $K3$. Thus, the perturbed ``ramified'' and ordinary invariants coincide up to a sign (see Theorem 3.3.2 of~\cite{nic}). In addition, we find that there are $h^{1,1} = 20$ admissible choices of embeddings  $D$ of a surface operator in $K3$.    

\bigskip\noindent{\it The Invariants on ${\bf T}^4$}

Let us consider  another (and the only other) closed hyper-K\"ahler manifold $X$ such as ${\bf T}^4$  with $b^{\pm}_2(X) > 1$ (and $R=0$) . Since $\chi(X) = 0$ and $\sigma(X) = 0$, we have ${x'}^2 = 2 \chi + 3 \sigma = 0$; that is, there is only one basic class for ${\bf T}^4$ -- the trivial one $x'_c = c_1(K)$. This is indeed consistent with the fact that $\textrm{dim}[H^0({\bf T}^4, K)] =1$, where $K$ is trivial. From our above discussion, we have\footnote{Note that even though $b_1({\bf T}^4) \neq 0$, since $\textrm{dim}[H^0({\bf T}^4, K)] =1$, the relevant moduli space is again made up of a single point; hence, one can still use the formula  ${\widetilde {SW}}(x'_c) = (-1)^w$ here.}  
\be
{\widetilde {SW}}_{{\bf T}^4} (0) = \pm 1.
\ee
Because  $b^+_2 (X) =3$, there are no wall-crossings, and the above result holds for all metrics on ${\bf T}^4$. Hence, the perturbed ``ramified'' and ordinary invariants coincide up to a sign  (see Theorem 3.3.2 of~\cite{nic}).  In addition, we find that there are $h^{1,1} = 4$ admissible choices of embeddings  $D$ of a surface operator in ${\bf T}^4$.

\bigskip\noindent{\it The Invariants on ${\bf P}^2$}

Now consider a K\"ahler manifold $X$ with $b^+_2(X) =1$ such as ${\bf P}^2$.  Since $h^{1,1}(X) =1$, there is a unique choice of embedding $D$ of the surface operator in ${\bf P}^2$. Because $X$ has $R > 0$, we have
 $SW_{{\bf P}^2} (x') = 0
 $ for some metric.\footnote{Although ${\bf P}^2$ is not spin, our earlier arguments relevant to the present analysis can be generalized to include non-spin manifolds.}  However, since $b^+_2(X) =1$, there can be wall-crossings. Nevertheless, since $b_1(X) =0$, the condition (\ref{wallcrossing condition}) for wall-crossings to occur is simply $b^-_2 > 9$. Since $b^-_2(X) < 9$, 
\be
SW_{{\bf P}^2} (x') = 0
\ee
for any metric on ${\bf P}^2$. Therefore, the \emph{unperturbed} ``ramified'' and ordinary invariants coincide in this case. 

\bigskip\noindent{\it The Invariants on ${\bf P}^2 \sharp N\overline {{\bf P}^2}$}

Let us consider $X$ to be the rational elliptic surface ${\bf P}^2 \sharp N\overline {{\bf P}^2}$
 given by the blow-up of ${\bf P}^2$ at $N$ points, where $N < 9$. Then, $b^+_2(X)$ =1 and $h^{1,1} = N+1$ -- hence, there are $N+1$ admissible choices of embeddings $D$ of the surface operator in $X$.  However, because $X$ has $R > 0$, 
$
SW_{{\bf P}^2 \sharp N \overline {{\bf P}^2}}(x') = 0
$
for some metric. Nevertheless, since $b^+_2(X) =1$, there can be wall-crossings. However, as $b_1(X) =0$, the condition (\ref{wallcrossing condition}) for wall-crossings to occur is simply $b^-_2 > 9$. Since $b^-_2(X) < 9$, 
\be
SW_{{\bf P}^2 \sharp N\overline {{\bf P}^2}}(x') = 0
\ee   
for any metric on ${\bf P}^2 \sharp N\overline {{\bf P}^2}$. Thus, the\emph{ unperturbed} ``ramified'' and ordinary invariants coincide.

\bigskip\noindent{\it The Invariants on ${\bf S}^2 \times \Sigma_g$}

Now consider $X$ to be a general product ruled surface ${\bf S}^2 \times \Sigma_g$ for $g > 0$. Then $b_1(X) = 2g$, $b^+_2(X) = b^-_2(X) =1$, and $\chi(X) = 4(1-g)$. In addition, we have ${\bf S}^2 \cap {\bf S}^2 = \Sigma_g \cap \Sigma_g = 0$, and $\Sigma_g \cap {\bf S}^2 = 1$.  Since $R > 0$  (for a metric where ${\bf S}^2$ small enough), we have
$
{\widetilde {SW}}_{{\bf S}^2 \times \Sigma_g} (x') = 0
$
for some pair $(\bar g, \rho_+)$. Nonetheless, there can be wall-crossings as explained, and ${\widetilde {SW}}_{{\bf S}^2 \times \Sigma_g} (x')$ will jump as we cross a ``wall'' into an adjacent chamber in the space of pairs.

Notice  that there are $h^{1,1} =2$ admissible choices of $D$; it can either be $\Sigma_g$ or ${\bf S}^2$. If $D = {\bf S}^2$, from (\ref{jump ruled}), the perturbed ``ramified'' SW invariants will, at some pair $({\bar g}', \rho'_+)$ in an adjacent chamber in the space of pairs, be given by
\be
SW_{{\bf S}^2 \times \Sigma_g} (x') = \pm[{l}]^g,
\label{eqn1}
\ee
as ${\bf S}^2 \cap {\bf S}^2 = 0$. On the other hand, if $D = \Sigma_g$, we have
\be
SW_{{\bf S}^2 \times \Sigma_g} (x') = \pm[p]^g,
\label{eqn2}
\ee
for some integer $p$, since $F' / 2\pi \in H^2(X, \mathbb Z)$. 

Notice that the dependence on the self-intersection number $D \cap D$ is not manifest in the above formulas; this is because for the above choices of $D$, we have $D \cap D =0$. Nevertheless, since the intersection form matrix of a four-manifold $X$ is  real, symmetric and unimodular, one can always diagonalize it using an orthogonal matrix; in other words, one can always find a basis of homology two-cycles $\{ U_i\}_{i =1, \dots, b_2 (X)}$ in (the torsion-free part of) $H_2(X, \mathbb Z)$, such that $U_i \cap U_j = n_i  \delta_{ij}$ for \emph{non-zero} integers $n_i$, whereby $\Pi_i n_i = \pm 1$;  consequently, there are $b_2(X)$ possible choices for $D$ with $D \cap D \neq 0$ -- namely, $D_i = U_i$ for $i =1, \dots, b_2(X)$.

In this case, the $h^{1,1} = b_2 =2$ admissible choices $D_1$ and $D_2$ will be given by $D_1 = a \cdot  {\bf S}^2 + b \cdot \Sigma_g$ and $D_2 = c \cdot {\bf S}^2 + d \cdot \Sigma_g$ for non-zero real numbers $a, b, c, d$, such that $ad = -bc$; $2ab = \pm 1$; $2cd = \pm 1$; and $D_i \cap D_j = n_i\delta_{ij}$ for integers $n_i$ obeying $n_1\cdot n_2 = \pm 1$. Also, one can express ${\bf S}^2 = m \cdot D_1 + n \cdot D_2$ for non-zero real numbers $m,n$, such that $ma = 1 - nc$ and $mb = -nd$. Hence, from (\ref{jump ruled}), the perturbed ``ramified'' SW invariants with nontrivially-embedded surface operators $D_1$ and $D_2$ will, at some pair $({\bar g}', \rho'_+)$ in an adjacent chamber in the space of pairs,  be given by
\be
{\widetilde {SW}}^{i}_{{\bf S}^2 \times \Sigma_g} (x') = \pm \left( {1 \over q_i}[{{l}_i} + r_i [c_1(L)] (\Sigma_g)] -  k_i   \alpha_i  (D_i \cap D_i)  \right)^g,
\label{eqn3}
\ee
where $i = 1, 2$,  $\{q_1, q_2 \}= \{a, c\}$, $\{ r_1, r_2 \}= \{b, d\}$, $\{ k_1, k_2 \}= \{m, n\}$, ${l}_i = \int_{D_i} {F / 2 \pi}$, and $[c_1(L)] (\Sigma_g) = -\int_{\Sigma_g} {F / 2\pi}$. 

The explicit dependence of the perturbed ``ramified'' SW invariants on the monopole numbers ${l}_i$, holonomy parameters $\alpha_i$ and self-intersection numbers $D_i \cap D_i$ of the two admissible surface operators, are manifest in the above formula. Clearly, from (\ref{eqn1}), (\ref{eqn2}) and (\ref{eqn3}), we see that the \emph{perturbed} ``ramified'' and ordinary invariants $\textrm{\it do not}$ necessarily coincide on a general product ruled surface with $g > 0$.

\vspace{1.0cm}
\hspace{-1.0cm}{\large \bf Acknowledgements:}\\
\vspace{-0.5cm}

I would first and foremost like to thank C.~LeBrun for his generous time and effort in educating me on various issues related to this work, and for his comments on a preliminary draft of this paper. I would also like to thank T.J.~Li and M.~Marcolli for useful email exchanges. This work is supported by the California Institute of Technology and the NUS-Overseas Postdoctoral Fellowship.


\vspace{0.0cm}


\begin{thebibliography}{99}

\bibitem{Moore-Witten}

G.~Moore and E.~Witten, ``Integration Over the $u$-plane in Donaldson Theory'', Adv.Theor.Math.Phys.{\bf 1}:298-387,1998. [arXiv:hep-th/9709193].

\bibitem{hirz}

F.~Hirzebruch and H.~Hopf, ``Felder von Flachenelementen in 4-dimensionalen Mannigfaltigkeiten,'' Math. Annalen {\bf 136} (1958) 156.

\bibitem{S-duality}

E.~Witten, ``On S-Duality in Abelian Gauge Theory'', Selecta Math.1:383,1995. [arXiv:hep-th/9505186].


\bibitem{mine}

M.C.~Tan, ``Surface Operators in N = 2 Abelian Gauge Theory'', 	JHEP 0909:047,2009. [arXiv:0906.2413].





\bibitem{monopoles}

E.~Witten, ``Monopoles and Four-Manifolds'', Math.Res.Lett. {\bf 1}:769-796,1994. [arXiv:hep-th/9411102].



\bibitem{Gukov-Witten}

S.~Gukov and E.~Witten, ``Gauge Theory, Ramification, And The Geometric Langlands Program'',  Current Developments in Mathematics Volume 2006 (2008), 35-180. [arXiv:hep-th/0612073]. 




\bibitem{QFT2}

``Quantum Fields and Strings, A Course for Mathematicians. Vol. 2'', AMS IAS. 

\bibitem{KM1}

P.B.~Kronheimer and T.S.~Mrowka, ``Gauge Theory for Embedded Surfaces: I'', Topology Vol. {\bf 32} (1993).

\bibitem{Marcos}

J.~Labastida and M.~Marino, ``Topological Quantum Field Theory and Four-Manifolds'',  Mathematical Physics Studies, Vol. {\bf 25}, Springer.  



\bibitem{Scorpan}

A.~Scorpan, ``The Wild World of 4-Manifolds'', AMS.





\bibitem{Donaldson}

S.~Donaldson, ``Irrationality and the $h$-Cobordism Conjecture'', J.~Diff.~Geom.~{\bf{26}} (1987) 141.  

\bibitem{Morgan}

John W.~Morgan, ``The Seiberg-Witten Equations and Applications to the Topology of Smooth Four-Manifolds'',  Princeton University Press.

\bibitem{le brun}

C.~LeBrun, ``Scalar-Flat K\"ahler Metrics On Blown-Up Ruled Surfaces,'' J.~Reine~Angew~Math. {\bf 420} (1991) 161. 

\bibitem{J.Moore}

J.D.~Moore, ``Lectures on Seiberg-Witten Invariants'', Lecture notes in Mathematics {\bf 1629}, Springer.  



\bibitem{Li}

T.J.~Li and A.K.~Liu, ``General Wall-Crossing Formula'', Math. Res. Lett. {\bf 2}, 797Ð810 (1995). 


\bibitem{nic}

L.I.~Nicolaescu, ``Notes on Seiberg-Witten Theory''. Graduate Studies in Mathematics, vol.~{\bf 28}, AMS. 





\end{thebibliography}
\end{document}